# High-mobility transport anisotropy and linear dichroism in few-layer black phosphorus

Jingsi Qiao,[1,2,§] Xianghua Kong,[1,2,§] Zhi-Xin Hu,[1,2] Feng Yang[1,2,3] and Wei Ji[1,2,*]

[1]*Department of Physics, Renmin University of China, Beijing 100872, China*

[2]*Beijing Key Laboratory of Optoelectronic Functional Materials & Micro-nano Devices,*
*Renmin University of China, Beijing 100872, China*

[3]*College of Physics and Electronic Engineering, Institute of Solid State Physics,*
*Sichuan Normal University, Chengdu 610068, China*

**Two-dimensional crystals are emerging materials for nanoelectronics. Development of the field requires candidate systems with both a high carrier mobility and, in contrast to graphene, a sufficiently large electronic bandgap. Here we present a detailed theoretical investigation of the atomic and electronic structure of few-layer black phosphorus (BP) in order to predict its electrical and optical properties. This system has a direct bandgap, tunable from 1.51 eV for a monolayer to 0.59 eV for a 5-layer sample. We predict that the mobilities are hole-dominated, rather high and highly anisotropic. The monolayer is exceptional in having an extremely high hole-mobility (of order 10000 cm$^2$ V$^{-1}$ s$^{-1}$) and anomalous elastic properties which reverse the anisotropy. Light absorption spectra indicate linear dichroism between perpendicular in-plane directions, which allows optical determination of the crystalline orientation and optical activation of the anisotropic transport properties. These results make few-layer BP a promising candidate for future electronics.**

---

[§] These authors contributed equally to this work.
[*] wji@ruc.edu.cn, http://sim.phys.ruc.edu.cn

The discovery of graphene laid the foundations for many new areas of research. One of the most important is the intensive investigation of two-dimensional (2D) atomic-layer systems, including graphene itself[1-5], transition metal dichalcogenides (TMDs)[5-8], silicene[9,10] and germanane[10,11], as candidate materials for future electronics applications[12-19]. A high-performance device such as a field-effect transistor (FET) requires a moderate electronic band gap, a reasonably high carrier mobility of the channel material and excellent electrode-channel contacts.[4,5,8,13-19] Graphene offers extremely high mobilities, due to its very low carrier effective mass, and thus is considered to be a promising candidate for high-speed FET devices, but its intrinsic dispersion is gapless.[2-4,12-14] Despite extensive efforts following a wide variety of approaches to the problem of opening a gap in different graphene nanostructures, all devices to date have a relatively large "off" current and thus a low "on-off" ratio.[13] For this reason the emergence of monolayer TMDs has attracted substantial research interest, and in particular $MoS_2$ has recently been used to fabricate a FET.[15] Unlike graphene, monolayer $MoS_2$ is a direct bandgap semiconductor with a carrier mobility of approximately 200 $cm^2$ $V^{-1}$ $s^{-1}$,[15] improvable up to 500 $cm^2$ $V^{-1}$ $s^{-1}$,[8] a value which is not unreasonable for applications but remains orders of magnitude lower than that of graphene.[13,14] Thus the entire community working on atomic-layer transport continues to search for a 2D material which is semiconducting, preferably with a direct gap, has high carrier mobility and has the potential to form excellent contacts with known electrode materials.

Here we present the theoretical discovery of a new category of layered direct-bandgap semiconductor, few-layer black phosphorus (BP), with high mobility, high in-plane anisotropy and linear dichroism. BP is an allotrope of phosphorus, whose crystal structure is a strongly folded honeycomb sheet with "troughs" running along the *y*-axis (*b* direction). We demonstrate using density functional theory (DFT) calculations that few-layer BP systems, from the monolayer up to 5-layer structures, are thermally stable with an interlayer interaction energy of -0.46 eV and a bandgap that falls exponentially with the thickness. Carrier mobilities at room temperature are high, and furthermore exhibit strongly anisotropic behaviour. The BP monolayer has an exceptionally high hole mobility, which we predict to lie in the range from 10000 to 26000 $cm^2$ $V^{-1}$ $s^{-1}$, and which we explain by visualizing the wavefunctions. We find an explicit linear dichroism in the computed absorption spectra, whereby the positions of the lowest-energy absorption peaks for the two in-plane directions differ strongly. These results demonstrate that few-layer BP is a new category of 2D semiconductor with high potential for novel applications in nanoelectronics and optoelectronics.

## Results

**Geometric and electronic properties of bulk BP.** Because no experimental data are available for the atomic structure and electronic properties of few-layer BP, we begin by considering bulk BP to refine the accuracy of our theoretical predictions. We have performed DFT calculations using a number of different functionals in order to gauge which provide the best fit to experiment; a full list may be found in Supplementary Table 1 and we summarize our findings here. If one considers only the lattice geometry then the PBE-G06[20, 21] and optB86b-vdW[22] methods produce the best results. However, neither LDA-mBJ[23, 24] nor HSE06[25, 26] calculations based on these ``best" geometries can provide a satisfactory value for the bulk bandgap, as shown in Supplementary Fig. 1. To optimize both sets of properties simultaneously, we found that the optB88-vdW functional,[22, 27] combined with either the LDA-mBJ or HSE06 method to predict the electronic bandstructure, gave the best fits. Details of choosing the best-fit functional are available in Supplementary Methods. Figure 1**a** shows the fully relaxed atomic structure of bulk BP, where the equilibrium lattice constants obtained from optB88-vdW are only 1–2% larger than the experiment.[28-30] The accompanying Brillouin zone (BZ) and electronic bandstructures, labeled mBJ(optB88-vdW) and HSE06(optB88-vdW) are shown in Fig. 1**b** and **c**. Both combinations of methods predict that bulk BP is a semiconductor with a direct bandgap at the Z point of 0.31 eV (mBJ) or 0.36 eV (HSE06), both values being fully consistent with the experimental value of 0.31–0.35 eV.[31-34]

At the bandgap wavevector, which is the Z point, one valence band (VB) and one conduction band (CB) disperse rather strongly along the Z-Q and Z-G directions (Fig. 1**c**), indicating very small effective masses. A fit of these bands using the nearly-free electron model gives effective carrier masses along the Z-Q direction which are rather small and similar, namely 0.12 $m_0$ for electrons and 0.11 $m_0$ for holes, whereas these along Z-G are slightly larger, taking the respective values 0.15 $m_0$ and 0.30 $m_0$ (all masses taken from HSE06). Considerably larger values are found along Z-T'-A', where the carrier effective masses are 1.15 $m_0$ and 0.71 $m_0$, respectively. All of these results are fully consistent with the experimental values for bulk BP.[35] The corresponding mBJ results are within 0.04 $m_0$ of the HSE06 values along Z-Q, Z-G and Z-T'-A' except for the electron effective mass along Z-T'-A', which is larger than the HSE06 value by 0.13 $m_0$. Values for the direction Z-Q are similar to other high-mobility semiconductors, such as AlGaAs (0.1 $m_0$),[36] and four times smaller than those of MoS$_2$.[37]

**Geometric and electronic properties of few-layer BP**. The similarity between the bandstructures of bulk and monolayer BP,[38] and the low effective masses of bulk BP,[35] suggest that few-layer BP is likely to be a high-mobility, direct-bandgap 2D semiconductor. We have thus performed direct calculations of the geometric and electronic properties for few-layer BP systems. Table I summarizes the changes of geometrical properties as a function of the layer number from 1 to 5. The lattice parameter $a$ increases by 0.11 Å on passing from bulk to monolayer BP, whereas $b$ grows only by 0.02 Å. There is an abrupt reduction of $a$ between the monolayer and the bilayer, which we believe to be the primary consequence of interlayer interactions (0.46 eV) in the bilayer. The significant stretching of $a$ in few-layer BP systems is caused almost exclusively by an increase of the bond angle $\theta 1$ but not of the bond lengths.

Results from our bandstructure calculations for the five few-layer BP systems are shown in Fig. 2. In the monolayer, the original Z point of the bulk BZ folds back to the Gamma (G) point, so that the original Z-Q and Z-T'-A' directions of the bulk BZ project onto the G-X and G-Y directions of the monolayer, which correspond respectively to the $a$ and $b$ directions in real-space (Fig. 2**a** and **b**). Results obtained from the mBJ method are quantitatively the same as those obtained from HSE06. Supplementary Fig. 2 and Supplementary Table 2 show that the predicted bandgap is rather less sensitive than the unit-cell dimensions to the functional used for optimizing the atomic structures, with similar values emerging for all few-layer BP systems from the best-performing functionals. However, other physical quantities, notably the location of band maximum (minimum) and the carrier effective mass (and hence the mobility) are in fact very sensitive to the choice of functional, and hence we have focused primarily on our optB88-vdW results for the discussion of these. Monolayer BP is indeed a direct-bandgap semiconductor, as shown in Fig. 2**c**, with the gap value of 1.51 eV obtained at the G point. Phosphorus is a heavier element than carbon and therefore should have stronger spin-orbit coupling (SOC) in its 2D forms than graphene. We have considered full SOC effects in calculating the electronic bandstructure of bulk BP. Supplementary Fig. 3 shows that inclusion of SOC terms has no appreciable effect on the primary features of the bandstructure, indicating that phosphorus is "not sufficiently heavy" to cause any qualitative changes. Only a very small separation of the formerly four-fold degenerate bands, into two band pairs, can be found from X along N (or M for few-layer BP) to Y, which reaches a maximum of 23 meV around the N (M) point. Along G-X and G-Y, these bands are already separated by bandstructure effects and SOC causes no further splitting. That the spin-orbit splitting energy is so small is a consequence of the small effective nuclear charge ($Z_{\text{eff}}$) of the P atom and the weak variation of the charge gradient in an elemental system such as BP.

When two monolayers are combined to form a bilayer (Fig. 2**d** and **e**), the gap is reduced to 1.02 eV and two additional bands emerge around the gap at the G point. Together with the original VB and CB, we denote these bands as VB1, VB2, CB1 and CB2, as shown in Fig. 2**f**. In real space, the states from these four bands with $k$ close to G are extended throughout the bilayer, as shown in Fig. 2**g**. States VB1 and VB2 differ in the interlayer region although they share the same origin in terms of atomic orbitals. A clear bonding-like feature is visible in the interlayer region (marked by red rectangles) for VB1, which lies lower in energy, whereas VB2 shows an anti-bonding feature. Similar bonding and anti-bonding features are also found in CB1 and CB2, where they are observed not between the layers but across the troughs. These features indicate that wavefunction overlap, rather than van der Waals effects, plays the primary role in mediating the interlayer interaction, which explains the abrupt reduction of the lattice constant $a$ from monolayer to bilayer. This interlayer interaction introduces band dispersions of VBs and CBs in the direction normal to the layers, which leads to the reduction of the bandgap (by 0.5 eV) from monolayer to bilayer.

The thicker the few-layer BP system, the stronger is the interlayer interaction and thus the larger is the dispersion of VBs (CBs), resulting in a smaller bandgap. Thus the gap falls continuously on adding more layers to the bilayer, reaching 0.59 eV in 5-layer BP, as illustrated in Fig. 2**h**. We fitted these values by an exponential decay relation and found that the corresponding bulk gap obtained by extrapolation is 0.53 eV, a value 0.17 eV larger than the one we calculate for bulk BP. We ascribe this difference to the elongated lattice parameter $a$ in few-layer BP and results calculated with a series of constrained $a$ values confirm our expectation; these results imply that the bandgap is very sensitive to the lateral strain along the $a$ ($x$) direction and could be modulated by varying interatomic separations and angles. During the preparation of this manuscript, we became aware that a strain-induced bandgap modulation has recently been demonstrated theoretically.[39]

The small effective masses remain in all of the few-layer BP systems (Table II). For the G-X direction in the monolayer, equivalent to Z-Q in the bulk, the carrier effective masses are 0.15 $m_0$ (hole) and 0.17 $m_0$ (electron), only 0.04-0.05 $m_0$ larger than those in bulk BP. It is remarkable that in the G-Y direction, equivalent to Z-T'-A' in the bulk, the VB appears to be nearly flat close to the G point, with an effective mass of 6.35 $m_0$ (see Fig. 2**c**), nine times its bulk value of 0.71 $m_0$. By contrast, the effective mass for the CB is 1.12 $m_0$, very close to its bulk value of 1.15 $m_0$. Unlike the bandgap, only the hole effective mass along G-Y shows a strongly layer-dependent evolution, in that it decreases from 6.35 $m_0$ for the monolayer to 1.81 $m_0$ for the bilayer and eventually to 0.89 $m_0$ in the 5-layer system, close to the bulk value of 0.71 $m_0$.

**Carrier mobility.** The electronic properties of few-layer BP are governed to a large extent by the carrier mobilities, which are in turn strongly influenced (but not solely determined) by their effective masses. Thus we also provide theoretical predictions for the carrier mobilities, in both *x* and *y* directions, in few-layer BP systems by applying a standard 2D model as discussed under Methods below. We apply a phonon-limited scattering model, in which the primary mechanism limiting carrier mobility is scattering due to phonons. [36, 40-42] Two properties of the few-layer BP lattice, namely the deformation potential $E_1$ and the elastic modulus $C_{2D}$ in the propagation direction of the longitudinal acoustic wave, are then the most relevant factors determining the mobility.[43, 44] These quantities were calculated using the optB88-vdW functional and the data are presented in Table II. If few-layer BP is cut into ribbons, a one-dimensional model should be adopted for predicting the carrier mobilities. We present the results and discussion for a 10-nm-width BP ribbon in the Supplementary Discussion and Supplementary Table 3.

Our predicted mobilities for the five few-layer BP systems (Table II) are in general moderately large (hundreds to thousands of cm$^2$ V$^{-1}$ s$^{-1}$), moderately anisotropic and asymmetric between electrons and holes, with the holes being more mobile in both directions. Concerning directional anisotropy, electron and hole mobilities for the *x* direction are nearly twice the values along *y* for holes and four times as large for electrons. However, the monolayer is exceptional in every way. The electron mobility along *x* is nearly 14, rather than four, times the value along *y*, i.e. 1100–1140 cm$^2$ V$^{-1}$ s$^{-1}$ compared with approximately 80 cm$^2$ V$^{-1}$ s$^{-1}$, which is consistent with the trend of the other few-layer systems. To our great surprise, however, the hole mobility along *x* is 16–38 times smaller than that along *y*, i.e. 640–700 cm$^2$ V$^{-1}$ s$^{-1}$ versus 10000–26000 cm$^2$ V$^{-1}$ s$^{-1}$, making *y* the direction of higher hole conductivity. This extraordinarily large value for monolayer BP is a consequence of the extremely small deformation potential, $E_{1y}$ = 0.15±0.03 eV, and occurs despite the fact that the carriers are very heavy (6.35 $m_0$). This ``monolayer exception" should be valuable in applications such as the separation of electrons and holes.

The value of $E_{1y}$ for holes in a monolayer is very striking in that it is an order of magnitude smaller than typical values of $E_1$, which are 5.0 eV for graphene,[44] 3.9 eV for MoS$_2$,[37] and 3.7 eV for h-BN [40] and 7 eV for AlGaAs quantum wells[36]. In few-layer BP systems, $E_{1y}$ for holes has a conventional value 2.97±0.18 eV in a

5-layer structure, decreasing smoothly to 1.63±0.16 eV in a bilayer. These results can be explained rather well by the form of the VB wavefunctions shown in Fig. 2**g**, where interlayer and stacking-induced intra-layer overlap increases with layer number. These effects are completely absent in monolayer BP, making the VB wavefunction quite isolated along *y,* as it is also for the VB2 (anti-bonding) state of a bilayer highlighted by the green rectangle in Fig. 2**g**. Thus small structural deformations due to longitudinal phonon oscillations along *y* have almost no effect on this wavefunction and cause little change in its energy, resulting in the very small deformation potential. The situation is quite different for the CB (electron) wavefunction, which has $E_{1x}$ = 2.72±0.02 eV in the monolayer and 5.02±0.02 eV in the bilayer. The in-plane bonding and anti-bonding features of the CB wavefunctions in few-layer BP systems cause a substantial enhancement of the energy-level perturbations due to structural deformations along *y*. The monolayer exception aside, all $E_1$ values for CB states are rather large (5.0–7.7 eV), for both *x* and *y* directions, and are almost independent of the number of layers, a result which can be explained by the strong CB wavefunction overlap, especially along *y* (Fig. 2**g**). For the VB states, the potentials are somewhat smaller (2–3 eV), precisely because their wavefunctions have less overlap between different P atoms.

**Optical absorption spectra and linear dichroism.** We have also predicted the optical absorption spectra of few-layer BP systems by computing the dielectric function. Two absorption spectra are shown in Fig. 3**a** and 3**b** for light incident along the *z* direction and linearly polarized in the *x* and *y* directions, respectively; while the results for incident light polarized in the *z* direction may be found in Supplementary Fig. 4. These results demonstrate a strong linear dichroism: for a dielectric polarization in the *x* direction, the band edge of the first absorption peak is found at the bandgap and thus falls rapidly with the thickness of the sample (Fig. 3**a**). By contrast, with *y*-polarized light this peak is found at 3.14 eV in the monolayer and its position falls only slightly with thickness, remaining at 2.76 eV in bulk BP (Fig. 3**b**). From the symmetries of the wavefunctions in Fig. 2**g**, clearly the (odd) dipole operator can connects the VB and CB states for *x*-polarization, allowing the direct bandgap process, but this is symmetry-forbidden for *y*-polarization and the transition occurs between VB and CB states elsewhere in the Brillouin zone.

## Discussion

High carrier mobilities would be expected in BP systems due to the small effective masses of 0.1 to 0.2

$m_0$. However, our calculations reveal that the key issue is not the effective mass but the very small elastic modulus $C_{2D}$: all the BP systems we have studied are ``very soft'' in the $x$ direction, due to the fact that bond angles are changed easily (with little effect on bond lengths). When compared to other monolayer systems, the modulus of BP for the $x$ direction is 29.0 J m$^{-2}$, a value respectively 2, 10 and 11 times smaller than MoS$_2$,[37] h-BN,[37,40] and graphene.[44] These results indicate that much higher mobilities could be most likely achieved if few-layer BP systems could be modified to become stiffer, i.e. to have a larger elastic modulus for the $x$ direction, for example by external stress, doping or substrate confinement. However, possible changes of the deformation potential as a result of modifying the elastic moduli are not yet clear and require further investigation. The modulus for the $y$ direction is significantly larger than for $x$, but the higher carrier effective masses, result in carrier mobilities being smaller than those along $x$. Both the bandgap and the mobility depend strongly on the structural properties and this offers a number of possibilities for further control over the electrical and mechanical properties of few-layer BP.

Further, the linear dichroism we find allows an electronic determination of the sample orientation. We propose an experimental set-up to deduce the orientation of few-layer BP systems using optical spectroscopy, as illustrated in Fig. 3**c**. If the incident light is linearly polarized in a chosen orientation and is near-normally incident, the absorption spectrum should vary when the sample is rotated and such that the $a$ and $b$ directions can be identified by monitoring the absorption signal. Once the sample orientation is determined, it is far easier to fabricate electrodes or gates utilizing the highest-mobility direction of a few-layer BP sample in an FET-type device.

In summary, we have shown theoretically that few-layer BP is a novel category of 2D semiconductor offering a direct bandgap, high carrier mobility and high transport anisotropy, all of which are tunable by controlling the layer thickness. The bandgap decreases from 1.51 eV for a monolayer to 0.59 eV for 5-layer BP, a set of values fitting well into the gap between graphene nanoribbons and TMDCs. The anisotropy of electric and optical properties is another unique feature distinguishing BP from other 2D materials, such as silicene[9,10], h-BN[5,45] and germanane[10,11]. Higher conductivity is generally found in the direction oriented perpendicular to the troughs (the $x$ direction) and holes are usually more mobile than electrons. For holes, the mobility in the $x$ direction increases from 600 cm$^2$ V$^{-1}$ s$^{-1}$ for a monolayer to over 4000 cm$^2$ V$^{-1}$ s$^{-1}$ for 5-layer BP. We found that monolayer BP is a very special material, in which we predict an extremely high hole mobility of 10000–26000 cm$^2$ V$^{-1}$ s$^{-1}$. All of our mobility results can be understood rather well from the form of the real-space wavefunctions, which show a much stronger interlayer coupling in few-layer BP than in graphene or TMDCs. All these results make few-layer BP a very promising candidate material for future applications in electronics and optoelectronics.

Note added in proof: during the review process for this manuscript, we became aware that few-layer BP samples have very recently been exfoliated from bulk BP and fabricated into electronic devices[46, 47]. Measurements on these samples verify the high hole mobilities[46, 47] and transport anisotropy[46] we predict. Although preliminary optical characterizations have been performed[46, 47], linear dichroism has not yet been explored.

## Methods

**Density functional theory calculation.** Density functional theory calculations were performed using the generalized gradient approximation for the exchange-correlation potential, the projector augmented wave method,[48, 49] and a plane-wave basis set as implemented in the Vienna *ab-initio* simulation package.[50] The energy cutoff for the plane-wave basis was set to 500 eV for all calculations. Two *k*-meshes of 10×8×4 and 10×8×1 were adopted to sample the first Brillouin zone of the conventional unit cell of bulk and few-layer BP, and the mesh density of *k* points was kept fixed when calculating bandstructures using primitive cells. In optimizing the system geometry, van der Waals interactions were considered by the vdW-DF level with the optB88 exchange functional (optB88-vdW).[22, 27] The shape and volume of each supercell were optimized fully and all atoms in the supercell were allowed to relax until the residual force per atom was less than 0.001 eV·Å$^{-1}$. Electronic bandstructures were calculated by the modified Becke-Johnson (mBJ)[23, 24] and hybrid functional (HSE06)[25, 26] methods based on the atomic structures obtained from the full optimization by optB88-vdW. The charge gradient used in the mBJ calculations, $c = 1.1574$, was extracted from the bulk value.

**Carrier mobility calculation.** In 2D the carrier mobility is given by the expression

$$\mu_{2D} = \frac{e\hbar^3 C_{2D}}{k_B T m_e^* m_d (E_1^i)^2} \quad \text{40-42} \tag{1}$$

where $m_e^*$ is the effective mass in the transport direction and $m_d$ is the average effective mass determined by $m_d = \sqrt{m_x^* m_y^*}$. The term $E_1$ represents the deformation potential constant of the valence-band minimum (VBM) for hole or conduction-band maximum (CBM) for electron along the transport direction, defined by

$E_1^i = \Delta V_i / (\Delta l / l_0)$. Here $\Delta V_i$ is the energy change of the $i^{th}$ band under proper cell compression and dilatation (calculated using a step of 0.5%), $l_0$ is the lattice constant in the transport direction and $\Delta l$ is the deformation of $l_0$. The elastic modulus $C_{2D}$ of the longitudinal strain in the propagation directions (both $x$ and $y$) of the longitudinal acoustic wave is derived from $(E - E_0)/S_0 = C(\Delta l / l_0)^2 / 2$ where $E$ is the total energy and $S_0$ is the lattice volume at equilibrium for a 2D system. All structural properties in the calculation of carrier mobilities were obtained from optB88-vdW and properties related to the electronic structure were computed with the HSE06 functional. The temperature used for the mobility calculations was 300 K.

**Uncertainty of deformation potential.** In calculations of the electron and hole mobilities, the deformation potential is derived from linear fits to the respective energies of the CBM and the VBM as functions of the lattice dilation or compression. Typical fitting results for $E_{1y}$ (electron state) and $E_{1x}$ (hole state) are shown in Supplementary Figs. **5a-c**. The slope represents the deformation potential, which usually has an unavoidable uncertainty accounted for in the standard fitting error. Supplementary Fig. **5d** summarizes the relative errors for the deformation potentials of electrons and holes in few-layer BP for the $x$ and $y$ directions. For a given direction and type of carrier, the error is generally smallest for monolayer BP, while that for 5-layer BP is the largest, except for the hole deformation potential along $y$. The error in $E_{1y}$ for hole states appears rather larger than the others, especially for monolayer BP where it exceeds 20%, presumably due to the small absolute value of $E_{1y}$ for this case. Supplementary Fig. **5c** shows details of the fit. The absolute value of $E_{1y}$ for hole states is only 0.15 eV, an order of magnitude smaller than the other energies; an error of 1 meV, believed to be the accuracy limit of DFT, may then lead to a fitting error of more than 10%. In this respect, a fitting error of 20% is definitely reasonable, and suggests that our calculation is highly accurate.

**Absorption spectra calculation.** Absorption spectra were calculated from the dielectric function using the expression $A(\omega) = 1 - e^{-\alpha(\omega) \cdot \Delta z}$, where $\alpha(\omega) = \dfrac{\omega \varepsilon_2}{cn}$ is the absorption coefficient, $n = \sqrt{\dfrac{\sqrt{\varepsilon_1^2 + \varepsilon_2^2} + \varepsilon_1}{2}}$ is the index of refraction, $\varepsilon_1$ and $\varepsilon_2$ are the real and imaginary parts of the dielectric function, $\omega$ is the light frequency, $c$ is the speed of light in vacuo and $\Delta z$ represents the unit-cell size in the $c$ direction. The electronic structures were obtained from HSE06 results and the $k$-mesh was doubled in calculating dielectric functions. Excitonic contributions were not considered in our calculations. The total number of bands considered was set to be twice that used in the total-energy and bandstructure calculations. Because the dielectric function is a tensor, the absorption spectra along the three directions $a$ ($x$), $b$ ($y$) and $c$ ($z$) were obtained separately.

**Acknowledgements**

We thank B. Normand for a critical reading of the manuscript. This work was supported by the National Natural Science Foundation of China (NSFC) under Grant Nos. 11004244 and 11274380, the Ministry of Science and Technology (MOST) of China under Grant No. 2012CB932704, the Beijing Natural Science Foundation (BNSF) under Grant No. 2112019 and the Basic Research Funds of Renmin University of China from the Central Government under Grant Nos. 12XNLJ03, 14XNH060 and 14XNH062. W.J. was supported by the Program for New Century Excellent Talents in Universities. Calculations were performed at the Physics Laboratory for High-Performance Computing of Renmin University of China and at the Shanghai Supercomputer Center.

**Author contributions**

W.J. conceived this research. J.Q., Z.X. H. and W.J. performed atomic and electronic structure calculations. X.K. calculated deformation potentials, elastic moduli, and carrier mobilities. F.Y. and W.J. predicted the optical absorption spectra. J.Q., X.K. and W.J. wrote the manuscript and all authors commented on it.

**Additional information**

Supplementary information is available in the online version of the manuscript. Reprints and permissions information is available online at www.nature.com/reprints. Correspondence and requests for materials should be addressed to W.J.

**Competing financial interests:** the authors declare no competing financial interests.

**Figure Legends**

**Figure 1 | Lattice and electronic structures of bulk black phosphorus. a**, Crystal structure of bulk BP marked with coordinate axes ($x, y, z$), lattice vectors ($a, b, c$) and structural parameters ($R1, R2, \theta1$ and $\theta2$). **b**, Brillouin zone path of BP primitive cell. **c,** Electronic bandstructures for bulk BP calculated with the HSE06 functional (red solid line) and the mBJ potential (blue dashed line), together with fitted effective masses along the Z-T'-A', Z-Q and Z-G directions. At the right of the image, a zoomed-in plot shows the direct bandgap at Z. $E_{\text{VBM}}$ is the energy of valence-band maximum.

**Figure 2 | Electronic structures of few-layer BP. a**, **b**, Top view of the atomic structure of the monolayer and the associated Brillouin zone. **d, e**, Side views of the atomic structure of the bilayer. **c**, **f**, Bandstructures of monolayer and bilayer BP calculated with the HSE06 functional (red solid lines) and the mBJ potential (blue dashed lines), respectively. Two valence (VB1 and VB2) and two conduction states (CB1 and CB2) are marked in panel **f**. **g**, Spatial structure of wavefunctions for the four marked states illustrated in the $xz$ and $yz$ planes using an isosurface of 0.0025 $e$ Å$^{-3}$. **h**, Evolution of the direct bandgaps as a function of the sample thickness. Functionals used for structural optimization are shown in parentheses. Bandgap values are marked for the monolayer system, for the extrapolation of our results and for real bulk BP.

**Figure 3 | Optical absorption spectra. a,b**, Optical absorption spectra of few-layer BP for light incident in the $c$ ($z$) direction and polarized along the $a$ ($x$) and $b$ ($y$) directions, respectively. Black dashed lines show a approximate linear fits used to estimate the band edges for the first absorption peak, which are highly anisotropic between $x$ and $y$. The band edge drops rapidly with sample thickness for $x$ [from 1.55 eV for the monolayer (red) to 0.60 eV for 5 layers (orange)] but only slightly for $y$ [from 3.14 eV to 2.90 eV]. **c**, Schematic illustration of a proposed experimental geometry to determine the orientation of few-layer BP structures using optical absorption spectroscopy, and thus to utilize the anisotropic electronic properties of BP. The light is linearly polarized in a chosen orientation and near-normally incident on the sample. The sample should be rotated to identify the $a$ or $b$ direction by monitoring the absorption signal, after which the source and drain electrodes, denoted as ``S'' and ``D'' respectively, may be deposited to fabricate an FET device.

## Tables

**Table I | Structural information for few-layer BP.**

| $N_L$ | $a$ (Å) | $b$ (Å) | $\Delta c$ (Å) | $R1$ (Å) | $R2$ (Å) | $\theta 1/\theta 1'$ (°) | $\theta 2/\theta 2'$ (°) |
|---|---|---|---|---|---|---|---|
| 1 | 4.58 | 3.32 | 3.20 | 2.28 | 2.24 | 103.51 | 96 |
| 2 | 4.52 | 3.33 | 3.20 | 2.28 | 2.24 | 102.96 | 96.21/95.92 |
| 3 | 4.51 | 3.33 | 3.20 | 2.28 | 2.24 | 102.81/102.74 | 96.30/95.99 |
| 4 | 4.50 | 3.34 | 3.20/3.21 | 2.28 | 2.24 | 102.76/102.67 | 96.34/96.01 |
| 5 | 4.49 | 3.34 | 3.20/3.21 | 2.28 | 2.24 | 102.71/102.63 | 96.37/96.05 |
| Bulk | 4.47 | 3.34 | 3.20 | 2.28 | 2.25 | 102.42 | 96.16 |

Lattice constants $a$, $b$ and $\Delta c$ (the interlayer spacing between two adjacent BP layers) and structural parameters $R1$, $R2$, $\theta 1$ and $\theta 2$ of few-layer and bulk BP calculated using optB88-vdW. There are slight differences in $\theta 1$ and $\theta 2$ between the outermost and inner layers.

**Table II | Predicted carrier mobility.**

| Carrier type | $N_L$ | $m_x^*/m_0$ G-X | $m_y^*/m_0$ G-Y | $E_{1x}$ (eV) | $E_{1y}$ (eV) | $C_{x\_2D}$ (J m$^{-2}$) | $C_{y\_2D}$ (J m$^{-2}$) | $\mu_{x\_2D}$ ($10^3$cm$^2$ V$^{-1}$ s$^{-1}$) | $\mu_{y\_2D}$ ($10^3$cm$^2$ V$^{-1}$ s$^{-1}$) |
|---|---|---|---|---|---|---|---|---|---|
| e | 1 | 0.17 | 1.12 | 2.72±0.02 | 7.11±0.02 | 28.94 | 101.60 | 1.10-1.14 | ~0.08 |
|  | 2 | 0.18 | 1.13 | 5.02±0.02 | 7.35±0.16 | 57.48 | 194.62 | ~0.60 | 0.14-0.16 |
|  | 3 | 0.16 | 1.15 | 5.85±0.09 | 7.63±0.18 | 85.86 | 287.20 | 0.76-0.80 | 0.20-0.22 |
|  | 4 | 0.16 | 1.16 | 5.92±0.18 | 7.58±0.13 | 114.66 | 379.58 | 0.96-1.08 | 0.26-0.30 |
|  | 5 | 0.15 | 1.18 | 5.79±0.22 | 7.35±0.26 | 146.58 | 479.82 | 1.36-1.58 | 0.36-0.40 |
| h | 1 | 0.15 | 6.35 | 2.50±0.06 | 0.15±0.03 | 28.94 | 101.60 | 0.64-0.70 | 10-26 |
|  | 2 | 0.15 | 1.81 | 2.45±0.05 | 1.63±0.16 | 57.48 | 194.62 | 2.6-2.8 | 1.3-2.2 |
|  | 3 | 0.15 | 1.12 | 2.49±0.12 | 2.24±0.18 | 85.86 | 287.20 | 4.4-5.2 | 2.2-3.2 |
|  | 4 | 0.14 | 0.97 | 3.16±0.12 | 2.79±0.13 | 114.66 | 379.58 | 4.4-5.2 | 2.6-3.2 |
|  | 5 | 0.14 | 0.89 | 3.40±0.25 | 2.97±0.18 | 146.58 | 479.82 | 4.8-6.4 | 3.0-4.6 |

Carrier types ``e'' and ``h'' denote ``electron'' and ``hole'', respectively. $N_L$ represents the number of layers, $m_x^*$ and $m_y^*$ are carrier effective masses for directions $x$ and $y$, $E_{1x}$ ($E_{1y}$) and $C_{x\_2D}$ ($C_{y\_2D}$) are the deformation potential and 2D elastic modulus for the $x$ ($y$) direction. Mobilities $\mu_{x\_2D}$ and $\mu_{y\_2D}$ were calculated using equation (1) with the temperature $T$ set to 300 K.

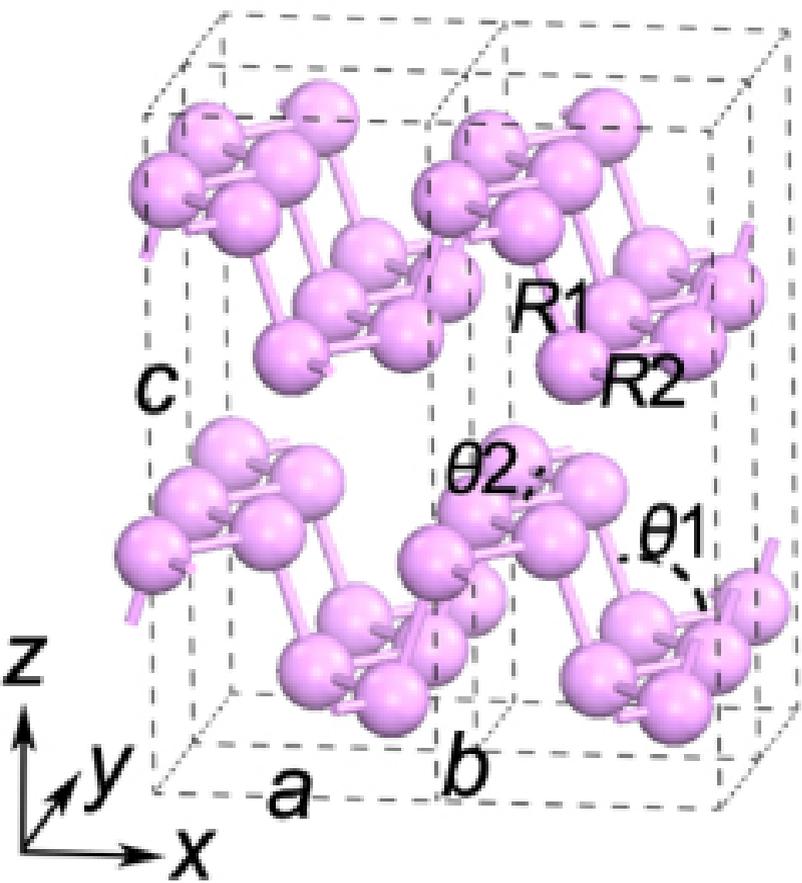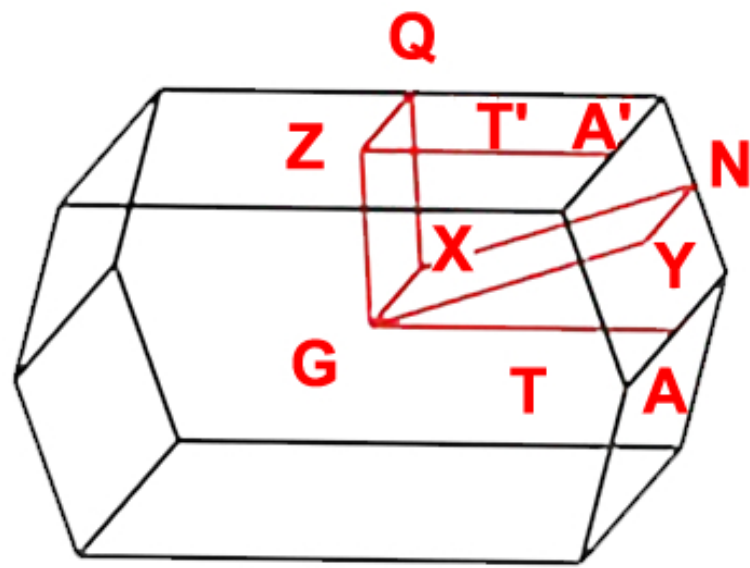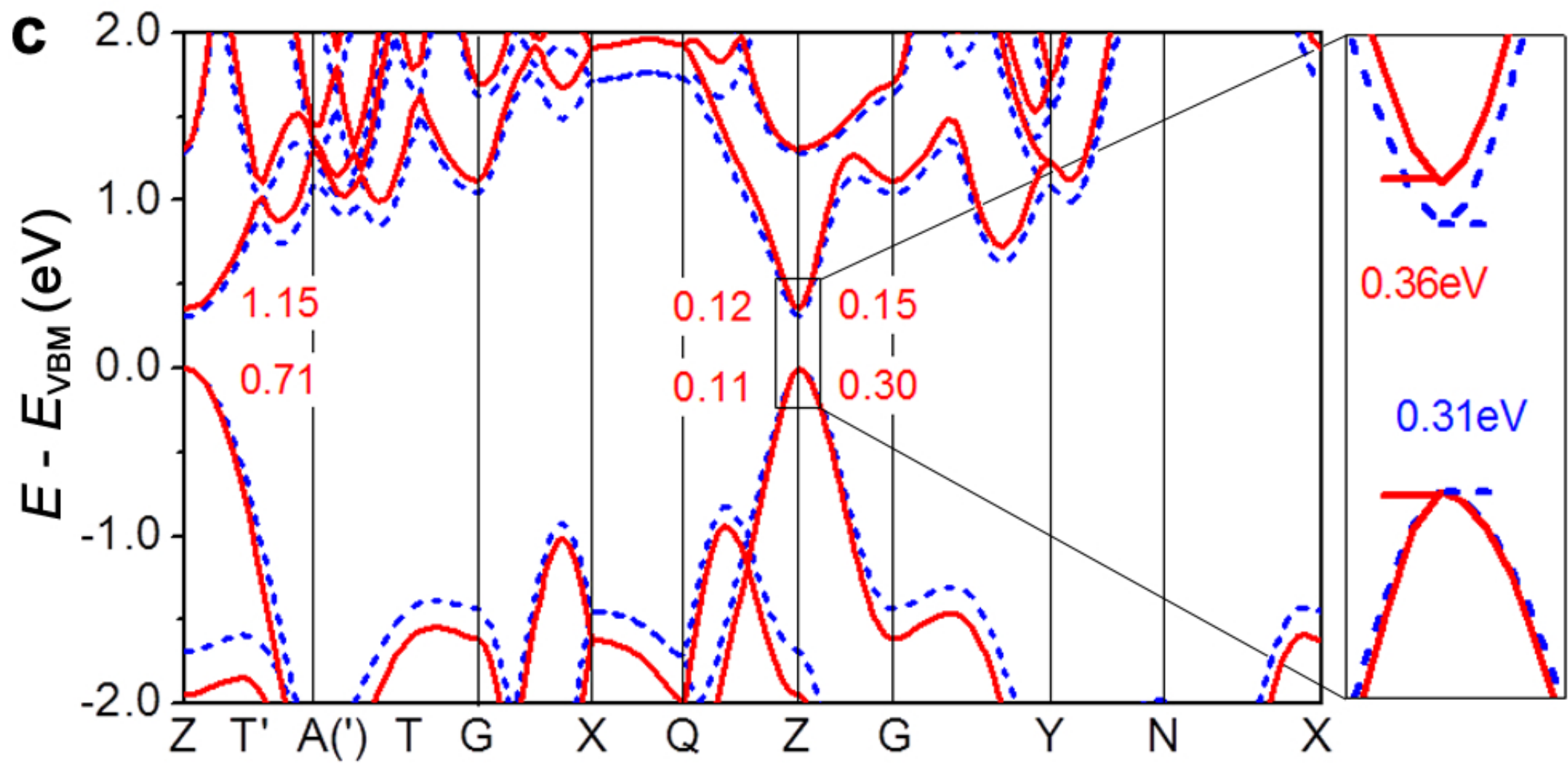

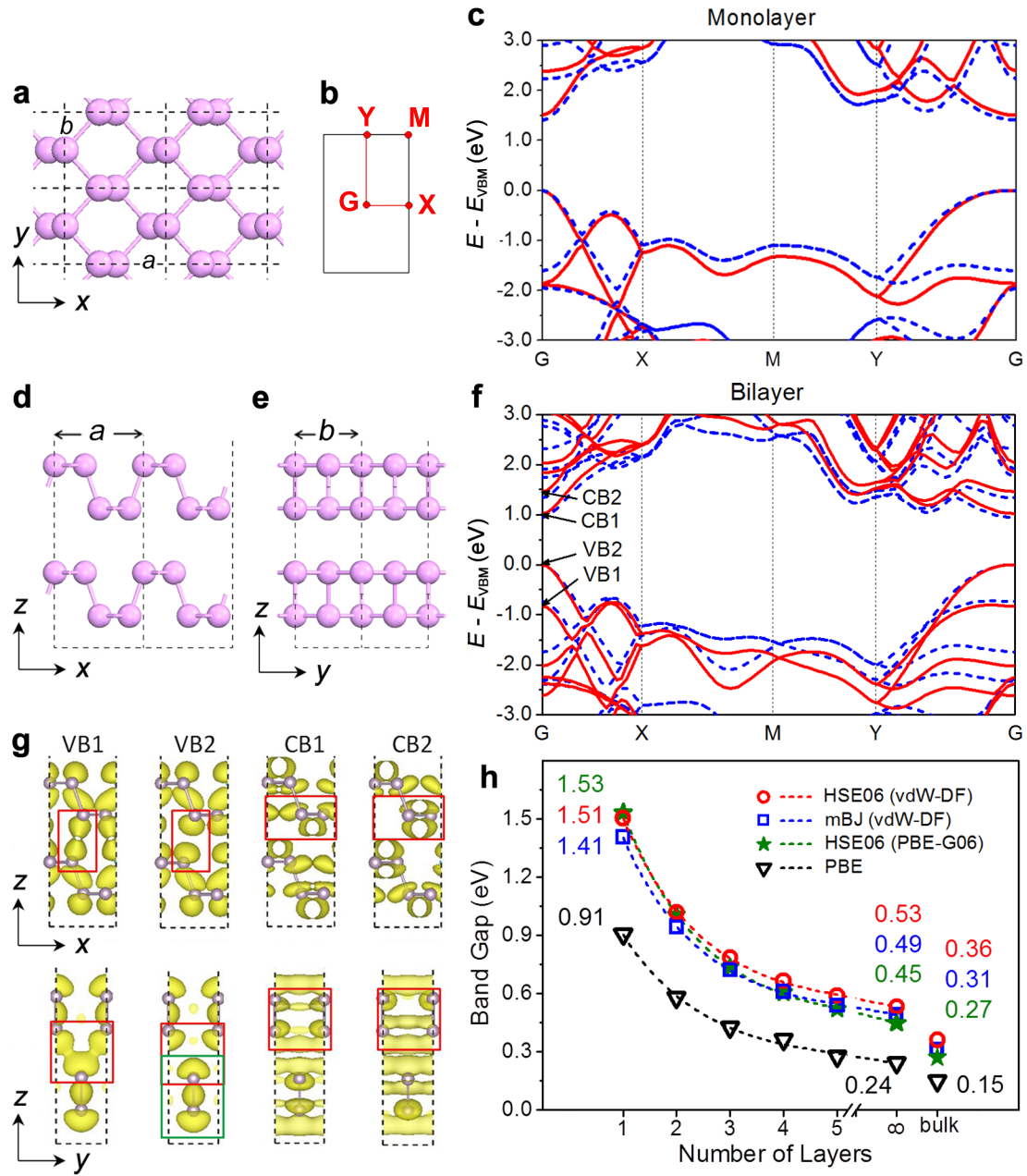

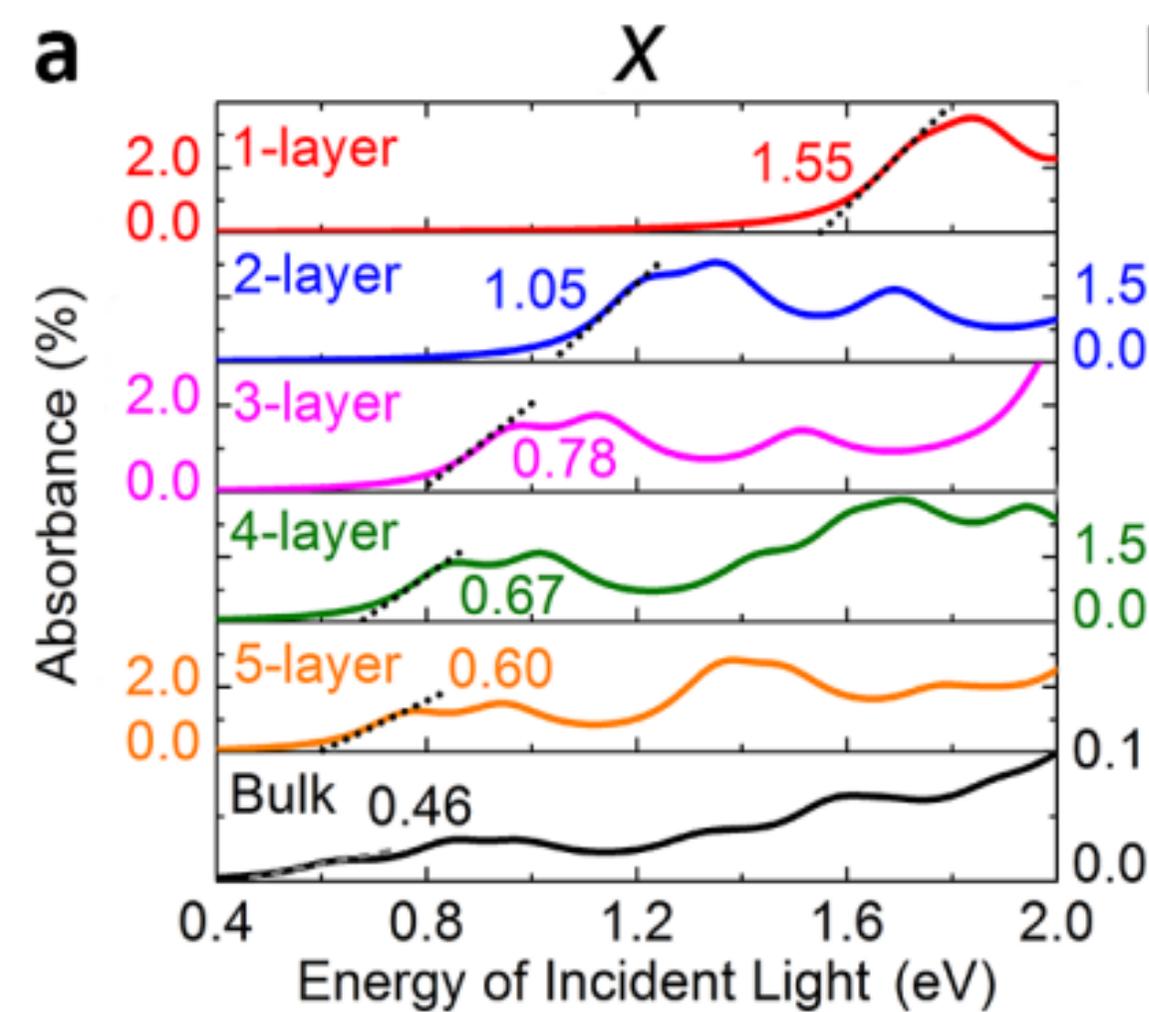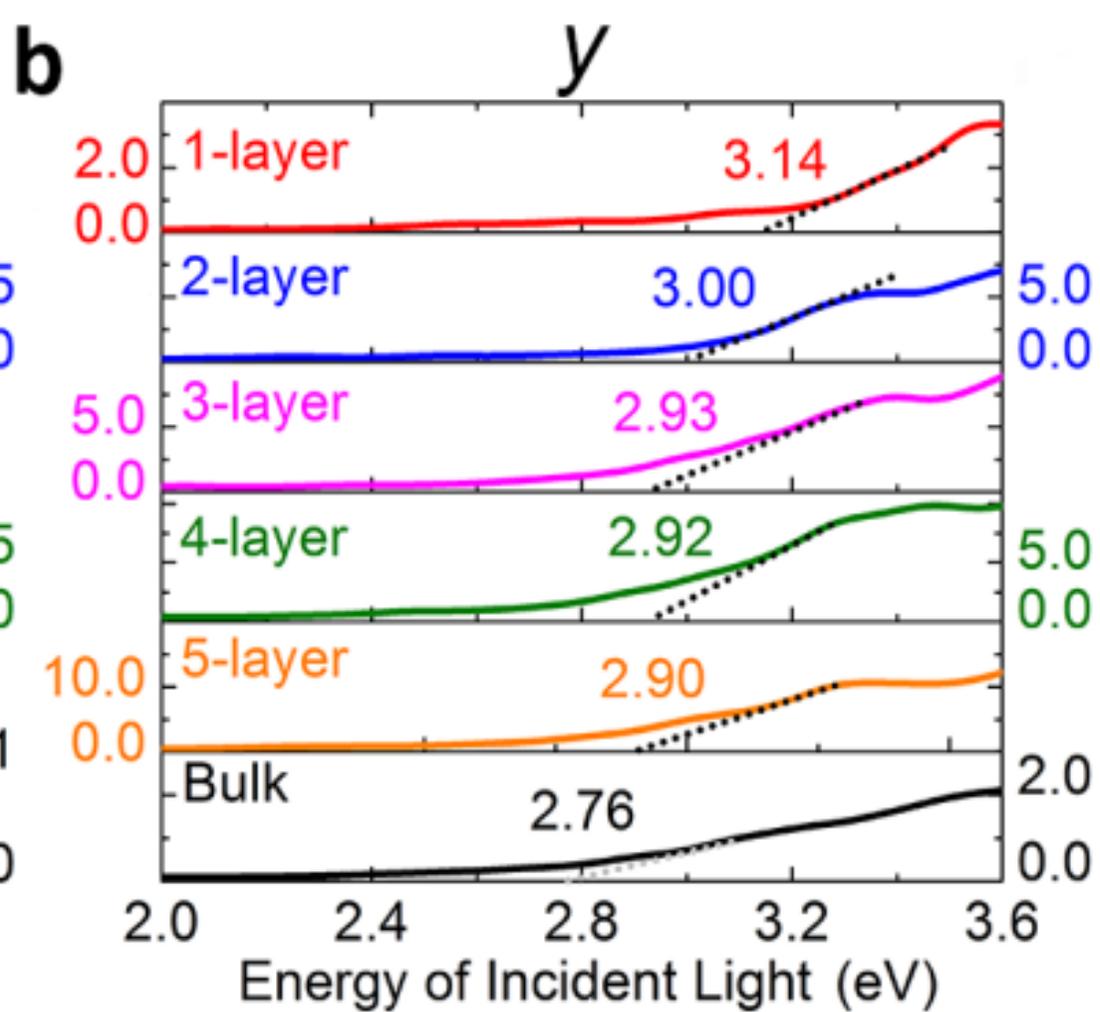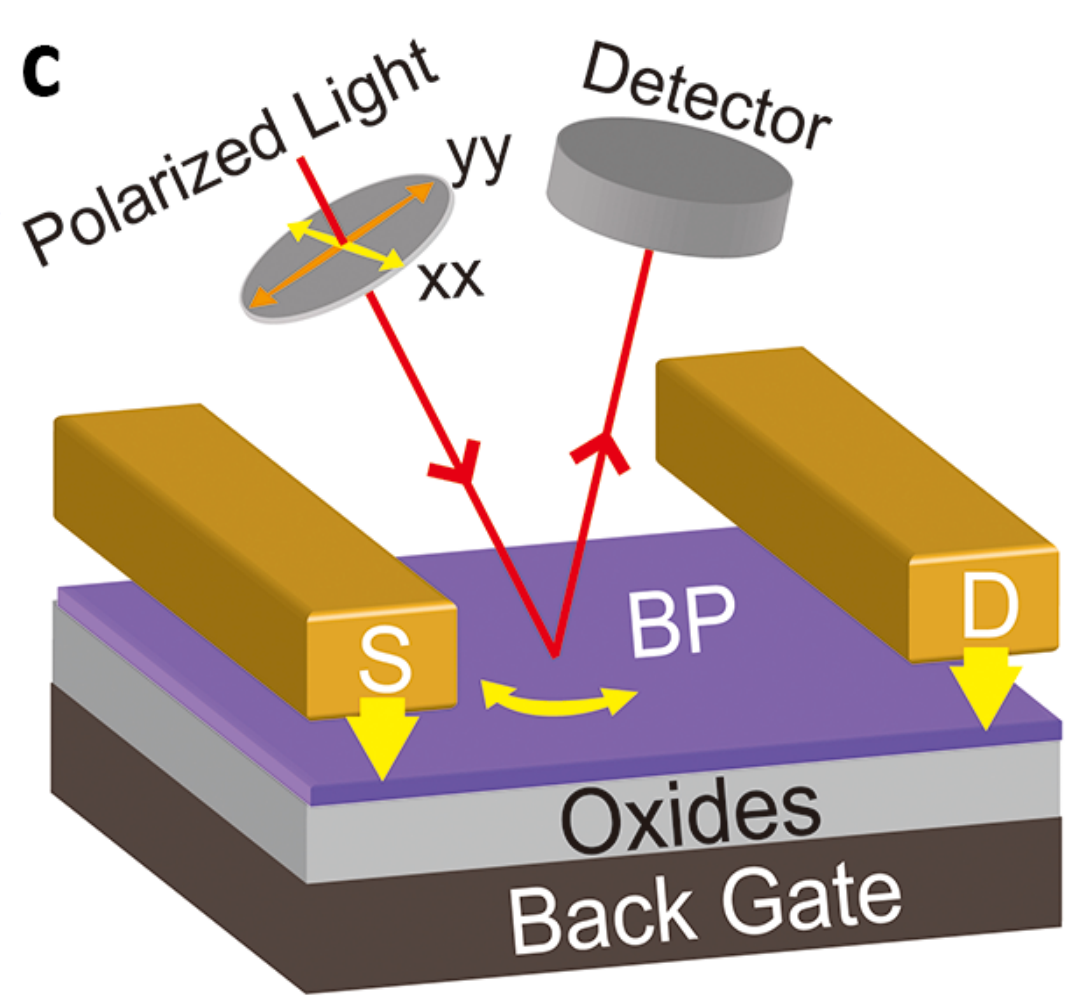

**Supplementary Figures**

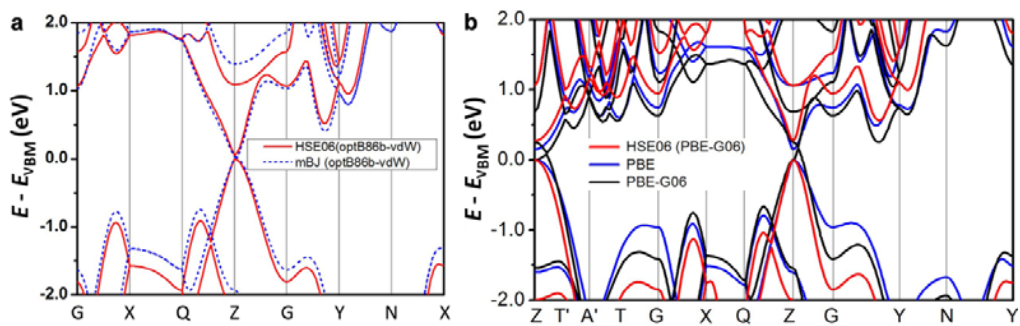

**Supplementary Figure 1 | Electronic bandstructures of bulk BP calculated with different functionals**. **a,** HSE06[1,2] (red solid lines) and LDA-mBJ[3,4] (blue dashed lines) bandstructures based on the atomic structure optimized by optB86b-vdW[5,6], denoted mBJ (optB86b-vdW) and HSE06 (optB86b-vdW). **b,** Bandstructures calculated with HSE06 (PBE-G06[7,8]) (red), PBE[7] (blue) and PBE-G06 (black).

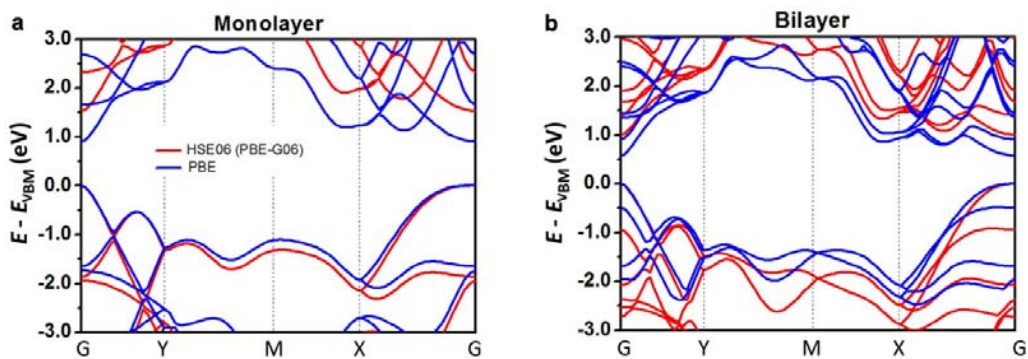

**Supplementary Figure 2 | Electronic bandstructures of monolayer, a, and bilayer, b, BP**. Results were obtained using the HSE06 functional based on the atomic structure from PBE-G06 (red) and PBE functional (blue).

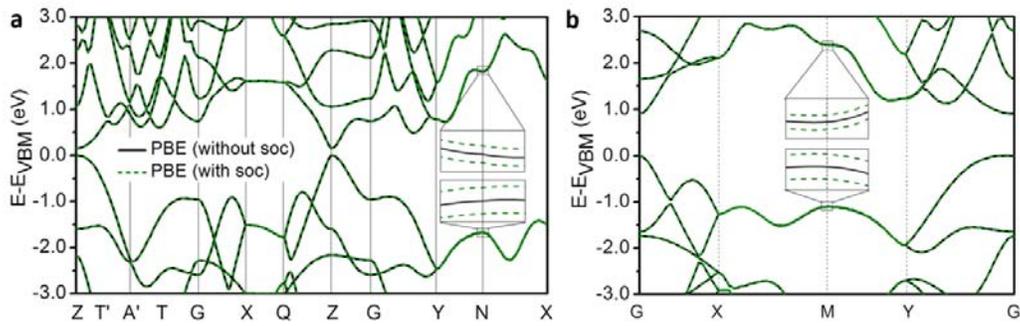

**Supplementary Figure 3 | Electronic bandstructures of bulk, a, and monolayer, b, BP, calculated by PBE with (green dashed lines) and without (black solid lines) SOC.** Insets provide a magnification to show SOC effects around the N (M) point.

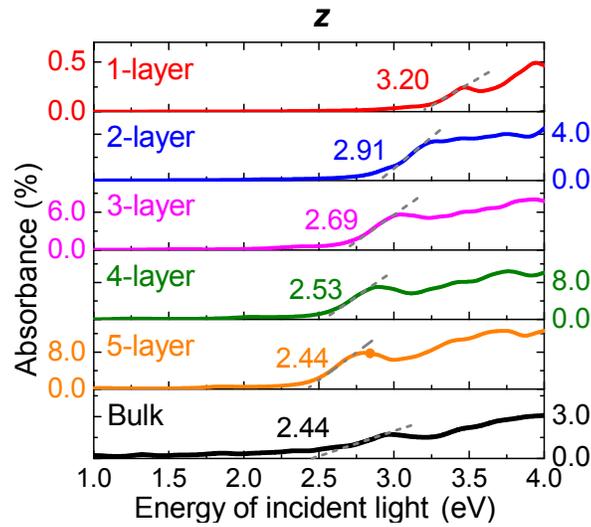

**Supplementary Figure 4 | Theoretically predicted optical absorption spectra of few-layer BP systems for light incident along an in-plane direction and polarized in *c (z)* direction.** Black dashed lines indicate the linear fit used to estimate the absorption edge. No appreciable absorption is found for energies below 2.0 eV. As for the case of *y*-polarized light shown in Fig. 3**b**, the bandgap process is symmetry-forbidden and absorption processes begin only beyond 2 eV. Thus the absorption exhibits an explicit anisotropy for light polarized in all three polarization directions. This behaviour should be readily detectable in spectroscopic measurements and may be exploited for optoelectronic applications.

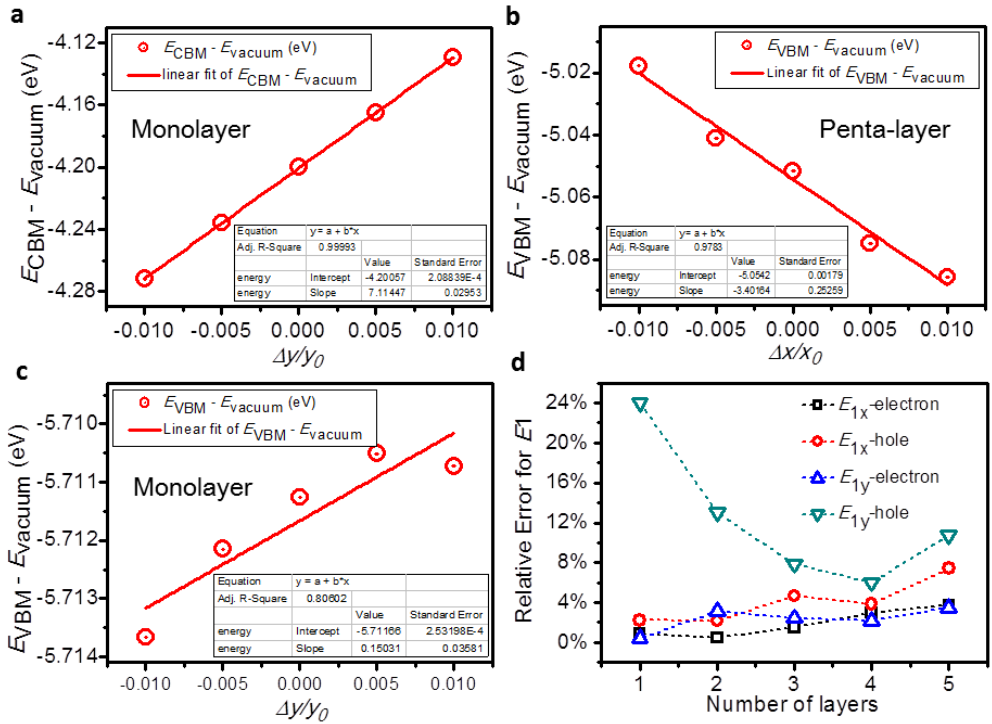

**Supplementary Figure 5 | Relative error in the deformation potential.** Band energy of the CBM of monolayer BP **a** and of the VBMs of 5-layer, **b,** and monolayer, **c,** BP with respect to the vacuum energy as a function of lattice dilation. Band energies were calculated with the HSE06 functional. Red solid lines are the fitting curves. Insets show the standard errors of the fitted slope, which corresponds to the deformation potential. All relative errors for the deformation potential of few-layer BP for the *x* and *y* directions are summarized in panel **d**.

**Supplementary Tables**

| Supplementary Table 1 \| Lattice constants and structural parameters of bulk black phosphorus calculated using different computational methods. | | | | | | | |
|---|---|---|---|---|---|---|---|
| Functional | $a$ (Å) | $b$ (Å) | $c$ (Å) | $R1$ (Å) | $R2$ (Å) | $\theta 1$ (°) | $\theta 2$ (°) |
| Expt.[9] | 4.38 | 3.31 | 10.48 | 2.24 | 2.22 | 102.09 | 96.34 |
| PBE | 4.57 | 3.30 | 11.33 | 2.26 | 2.22 | 103.59 | 95.98 |
| PBE[10] | 4.54 | 3.28 | 11.22 | - | - | - | - |
| PBE-G06 | 4.43 | 3.32 | 10.49 | 2.26 | 2.23 | 102.48 | 96.50 |
| PBE-G06[10] | 4.40 | 3.30 | 10.43 | - | - | - | - |
| PBE-TS[10] | 4.39 | 3.29 | 10.82 | - | - | - | - |
| RPBE-G06 | 4.57 | 3.33 | 10.91 | 2.27 | 2.23 | 103.58 | 96.42 |
| HSE06-G06 | 4.42 | 3.30 | 10.43 | 2.23 | 2.20 | 102.87 | 96.96 |
| optPBE-vdW | 4.54 | 3.34 | 10.95 | 2.28 | 2.24 | 103.02 | 96.07 |
| optB86b-vdW | 4.35 | 3.33 | 10.52 | 2.27 | 2.24 | 101.58 | 96.09 |
| optB88-vdW | 4.47 | 3.34 | 10.71 | 2.28 | 2.25 | 102.42 | 96.16 |
| PW91 (S1) | 4.55 | 3.31 | 11.18 | 2.27 | 2.23 | 103.45 | 96.04 |
| PW91 (S2) | 4.56 | 3.31 | 11.19 | 2.27 | 2.23 | 103.46 | 96.03 |
| PW91-G06 | 4.43 | 3.33 | 10.48 | 2.27 | 2.23 | 102.47 | 96.57 |
| PW91[11] | 4.42 | 3.35 | 10.59 | 2.26 | 2.24 | 102.31 | 96.85 |
| The functionals adopted are PBE, RPBE[12] and HSE06 and the dispersion forces are taken into account at the DFT-G06[8] and vdW-DF[13] levels. | | | | | | | |

| Supplementary Table 2 \| Bandgaps of few-layer BP. | | | | |
|---|---|---|---|---|
| Number of Layers | HSE06 (optB88-vdW) | mBJ (optB88-vdW) | HSE06 (PBE-G06) | PBE (PBE) |
| 1 | 1.51 | 1.41 | 1.53 | 0.91 |
| 2 | 1.02 | 0.94 | 1.01 | 0.58 |
| 3 | 0.79 | 0.72 | 0.73 | 0.42 |
| 4 | 0.67 | 0.61 | 0.61 | 0.36 |
| 5 | 0.59 | 0.54 | 0.52 | 0.28 |
| ∞ | 0.53 | 0.49 | 0.45 | 0.24 |
| Bulk | 0.36 | 0.31 | 0.27 | 0.15 |
| Exact values, in units of eV, of bandgaps plotted in Fig. 2**h** and Supplementary Fig. 2. | | | | |

| Carrier type | $N_L$ | $m_x^*/m_0$ G-X | $m_y^*/m_0$ G-Y | $E_{1x}$ (eV) | $E_{1y}$ | $C_{x\_1D}$ ($10^{-7}$ J m$^{-1}$) | $C_{y\_1D}$ | $\mu_{x\_1D}$ ($10^3$ cm$^2$ V$^{-1}$ s$^{-1}$) | $\mu_{y\_1D}$ |
|---|---|---|---|---|---|---|---|---|---|
| e | 1 | 0.17 | 1.12 | 2.72±0.02 | 7.11±0.02 | 2.90 | 10.16 | 5.4-5.6 | ~0.24 |
|   | 2 | 0.18 | 1.13 | 5.02±0.02 | 7.35±0.16 | 5.74 | 19.46 | ~3.0 | 0.38-0.44 |
|   | 3 | 0.16 | 1.15 | 5.85±0.09 | 7.63±0.18 | 8.58 | 28.72 | 3.8-4.0 | 0.52-0.58 |
|   | 4 | 0.16 | 1.16 | 5.92±0.18 | 7.58±0.13 | 11.46 | 37.94 | 4.8-5.4 | 0.70-0.76 |
|   | 5 | 0.15 | 1.18 | 5.79±0.22 | 7.35±0.26 | 14.66 | 47.98 | 6.8-8.0 | 0.88-1.0 |
| h | 1 | 0.15 | 6.35 | 2.50±0.06 | 0.15±0.03 | 2.90 | 10.16 | 7.6-8.2 | 26-66 |
|   | 2 | 0.15 | 1.81 | 2.45±0.05 | 1.63±0.16 | 5.74 | 19.46 | 16-17 | 3.2-5.4 |
|   | 3 | 0.15 | 1.12 | 2.49±0.12 | 2.24±0.18 | 8.58 | 28.72 | 22-26 | 5.6-7.8 |
|   | 4 | 0.14 | 0.97 | 3.16±0.12 | 2.79±0.13 | 11.46 | 37.94 | 20-24 | 6.2-7.8 |
|   | 5 | 0.14 | 0.89 | 3.40±0.25 | 2.97±0.18 | 14.66 | 47.98 | 20-28 | 7.2-11 |

**Supplementary Table 3 | Carrier mobility predicted by the 1D model.**

$E_{1x}$ ($E_{1y}$) and $C_{x\_1D}$ ($C_{y\_1D}$) are the deformation potential and effective elastic modulus of a 10 nm BP ribbon for the $x$ ($y$) direction. Mobilities $\mu_{x\_1D}$ and $\mu_{y\_1D}$ were calculated using Supplementary Equation (2).

## Supplementary Discussion

**Predicted carrier mobility of few-layer BP with other models**

Because the anisotropies of the carrier mobilities are crucial in determining the physical properties of few-layer BP systems, we discuss this issue in more detail here. First, an alternative expression for the mobility which can be found in a number of studies of anisotropic 2D systems is [14-17]

$$\mu_{x\_2D\_AS} = \frac{2e\hbar^3 C_{x\_2D}}{3k_B T (m_e^*)^2 E_{1x}^2} \quad (1)$$

where $m_e^*$ is the effective mass along the transport direction and $T$ is the temperature. We believe that this model, which contains no effective connection between the perpendicular directions, overestimates the extent of anisotropy and therefore we do not consider it further here. An expression for the mobility in 1D systems is [18]

$$\mu_{x\_1D} = \sqrt{\frac{2}{\pi}} \frac{e\hbar^2 C_{x\_1D}}{(k_B T)^{1/2} (m_e^*)^{3/2} E_{1x}^2} \quad (2)$$

which should be relevant for BP ribbons. The narrower the ribbon, the smaller is its effective elastic modulus, and in Supplementary Table 3 we present mobility values calculated from Supplementary Eq. (2) for BP ribbons of 10nm in width, which is a reasonable value for experimental fabrication. Our results show that the 1D mobilities are rather large compared with the values obtained in the 2D model used in the main text. Thus we suggest that etching a 2D BP sheet to produce a 1D ribbon could enhance the carrier mobilities quite significantly, although we caution that the edge structure and its role in electron scattering require further investigation.

## Supplementary Methods

**Best-fit functionals for few-layer black phosphorus**

Supplementary Figure 1 shows that methods HSE06-G06[1, 2, 8], PBE-G06[7, 8] and optB86b-vdW[5] reproduce very accurately the experimental geometry and the theoretical equilibrium volumes are respectively just 0.1% smaller, 1% larger and 0.1% larger than the experiment.[9, 19] The results using PBE[7] and PBE-G06 are consistent with a previous study[10]. In calculations for the exchange-correlation energy performed with PW91[20], we used two approaches to find the equilibrium crystal structure of bulk BP. The results from these two schemes are consistent with each other and, as expected, similar to those of PBE, but differ significantly from the PW91 results of one other study.[11] However, our PW91-G06 results do appear similar to the ``PW91'' results of that work [11]. The functional optB88-vdW[5, 6] provides the second best results in terms of geometry, yielding an equilibrium volume only 5% larger than the experiment. The calculated bond lengths are similar for all methods other than HSE06-G06, which predicts bond lengths considerably smaller than any of the others.

Although the geometry is very accurately reproduced by optB86b-vdW, the associated bandgaps are only 0.03 eV (mBJ) and 0.05 eV (HSE06). Similarly small gaps are obtained in the mBJ and HSE06 bandstructures based on the PBE-G06 or HSE06-G06 geometries. The PBE-G06 bandstructure even reports a negative band gap. Thus we conclude that the combinations of mBJ and HSE06 functionals with the PBE-G06, HSE06-G06 and optB86b-vdW schemes are not acceptable for predicting the electronic structure of few-layer BP. In addition, BP appears to be metallic, with a band crossing near the Z point in the PBE-G06 bandstructure; by contrast, in the PBE bandstructure one finds a direct gap of 0.15 eV, a result we ascribe to the fact that PBE overestimates the unit-cell volume, giving rise to a smaller overlap, a weaker interaction and thus a larger gap than that obtained by PBE-G06. The combination of HSE06 or mBJ with the optB88-vdW structure thus provides the best prediction for BP, offering the second-best geometry and the best electronic bandstructure

# **Supplementary References**